# Processing a Trillion Cells per Mouse Click

Alexander Hall, Olaf Bachmann, Robert Büssow, Silviu Gănceanu, Marc Nunkesser
Google, Inc.
{alexhall, olafb, buessow, silviu, marcnunkesser}@google.com

## ABSTRACT

Column-oriented database systems have been a real game changer for the industry in recent years. Highly tuned and performant systems have evolved that provide users with the possibility of answering ad hoc queries over large datasets in an interactive manner.

In this paper we present the column-oriented datastore developed as one of the central components of PowerDrill[1]. It combines the advantages of columnar data layout with other known techniques (such as using composite range partitions) and extensive algorithmic engineering on key data structures. The main goal of the latter being to reduce the main memory footprint and to increase the efficiency in processing typical user queries. In this combination we achieve large speed-ups. These enable a highly interactive Web UI where it is common that a single mouse click leads to processing a trillion values in the underlying dataset.

## 1. INTRODUCTION

In the last decade, large companies have been placing an ever increasing importance on mining their in-house databases; often recognizing them as one of their core assets. With this and with dataset sizes growing at an enormous pace, it comes as no surprise that the interest in column-oriented databases (*column-stores*) has grown equally. This spawned several dozens of research papers and at least a dozen of new column-store start-ups, cf. [2]. This is in addition to well established offerings, e.g., by MonetDB [25], Netezza [26], or QlikTech [30]. Since 2011 all major commercial database vendors actually provide column-store technologies (cf. [25]).

Typically, these products are deployed to import existing databases into the respective column-store. An OLAP or OLTP, i.e., SQL, interface is provided to then mine the data interactively. The key advantage shared by these systems is that column-oriented storage enables reading only data for relevant columns. Obviously, in denormalized datasets with often several thousands of columns this can make a huge difference compared to the the row-wise storage used by most database systems. Moreover, columnar formats compress very well, thus leading to less I/O and main memory usage.

[1]Internal project name only, following a Google tradition of choosing names of wood-processing tools for "logs" analysis.

At Google multiple frameworks have been developed to support data analysis at a very large scale. Best known and most widely used are MapReduce [13] and Dremel [23]. Both are highly distributed systems processing requests on thousands of machines. The latter is a column-store providing interactive query speeds for ad hoc SQL-like queries.

In this paper we present an alternative column-store developed at Google as part of the PowerDrill project. For typical user queries originating from an interactive Web UI (developed as part of the same project) it gives a performance boost of 10–100x compared to traditional column-stores which do full scans of the data.

**Background**

Before diving into the subject matter, we give a little background about the PowerDrill system and how it is used for data analysis at Google. Its most visible part is an interactive Web UI making heavy use of AJAX with the help of the Google Web Toolkit [16]. It enables data visualization and exploration with flexible drill down capabilities. In the background, the "engine" provides an abstraction layer for the UI based on SQL: the user constructs charts via drag'n'drop operations, they get translated to group-by SQL queries, which the engine parses and processes. It can send out such queries to different backends, e.g., Dremel, or execute them directly on data stored, e.g., in CSV files, record-io files (binary format based on protocol buffers [29]), or in Bigtable [10]. The third large part of the project is the column-store presented in this paper.

The Web UI is very versatile; it allows to select arbitrary dimensions, measures, and computed values for grouping and filtering. The dimensions can have a large number of distinct values, such as strings representing Google searches. A user can quickly drill down to values of interest, e.g., all German searches from yesterday afternoon that contain the word "auto", by restricting a set of charts to these values. For these reasons, pre-aggregation or indexing of data does not help and we need to query the raw data directly.

The nature of the use cases enabled by this UI demand for high availability and low latency. Examples of such use cases include: Responding to customer requests, spam analysis, dealing with alerts in highly critical revenue systems, or monitoring and assessing changes to production systems.

The system has been in production since end of 2008 and was made available for internal users across all of Google mid 2009. Each month it is used by more than 800 users sending out about 4 million SQL queries. After a hard day's work, one of our top users has spent over 6 hours in the UI, triggering up to 12 thousand queries. When using our column-store as a backend, this may amount to scanning as much as 525 trillion cells in (hypothetical) full scans.

The column-store developed as part of PowerDrill is tailored to support a few selected datasets and tuned for speed on typical queries resulting from users interacting with the UI. Compared to Dremel which supports thousands of different datasets (streaming the data from a distributed file system such as GFS [15]), our column-store relies on having as much data in memory as possible. PowerDrill can run interactive single queries over more rows than Dremel, however the total amount of data it can serve is much smaller, since data is kept mostly in memory, whereas Dremel uses a distributed file system.

This and several other important distinctions, enable handling very large amounts of data in interactive queries. Consider a typical use case such as triggering 20 SQL queries with a single mouse click in the UI. In our production system *on average* these queries process 782 billion cells in 30-40 seconds (under 2 seconds per query), several orders of magnitude faster than what a more traditional approach as used by Dremel could provide.

**Contributions**

The main contributions presented in this paper:

- We describe how—unlike in most column-stores—the data is partitioned and organized in an import phase (Section 2.2). This enables skipping and caching large parts of the data: on average in production 92.41% is skipped and 5.02% cached, leaving only 2.66% to be scanned (see also Section 6).

- We present the basic data-structures used in Section 2.3. Their main goal is to support the partitioned layout of the data and to enable quick skipping of chunks of data. For optimal usage it is assumed they can be held in memory.

  Experiments show that these simple data-structures also directly give performance benefits of around 100x or more on full scans, compared to two row-wise storage formats and Dremel's column-store (Section 2.5). Note that for these experiments we do *not* partition the data at import.

  When dropping the "in memory" assumption, a still impressive factor of 30x can be achieved.

- In Section 3 we present several successive "algorithmic engineering" choices to improve key data-structures. The aim being to reduce the memory footprint for certain typical cases. We pin-point the effects of individual optimizations with experiments measuring memory usage. E.g., for the important case of a field with many distinct values, we obtain a reduction of 16x.

- In Section 4 we describe how queries may be computed in a distributed manner on a cluster of machines. In Section 5 we present selected extensions and finally in Section 6 the highly distributed setup of the actual productionized system running on over 1000 machines.

We give measurements concerning the usage in practice which show the positive effect of the partitioning (enabling to skip or cache large parts of the data).

**Related Work**

For an introduction to OLAP and basic techniques applied in data-warehouse applications, see the Chaudhuri and Dayal [11].

To obtain an overview of recent work on column-store architectures, please see the concise review [2] and references therein. The excellent PhD thesis by Abadi [1] can serve as a more in-depth introduction and overview of the topic.

Recent research in this area includes, e.g., work on how super-scalar CPU architectures affect query processing [9], tuple reconstruction [17], compression in column-stores [34, 9, 3], and a comparison to traditional row-wise storage [4]. Kersten et al. [20] give a more open ended outlook on interesting future research directions.

The plethora of open-source and commercial column-store systems, e.g., [34, 25, 26, 30, 36] further demonstrates the effectiveness of this paradigm.

Melnik et al. [23] recently have introduced Dremel to a wider audience. As mentioned, its power lies in providing interactive responses to ad hoc SQL queries over thousands of datasets. It achieves this by streaming over petabytes of data (stored, e.g., on GFS [15]) in a highly distributed and efficient manner. This is also a key difference to the column-store presented in this paper which heavily relies on having as much data in memory as possible and therefore only is used for a few selected data sources. Melnik et al. also give a nice overview of data anlysis at Google and how interactive approaches like Dremel's complement the offline MapReduce [13] framework.

Skipping over data in the context of colum-stores has been explored by other authors, e.g., Slezak et al. [32] or Moerkotte [24]. We give some details on these approaches in comparison to ours in Section 2.1.

Reordering rows to improve the compression of column-wise stored data has been investigated, e.g., by [18, 21, 3]. We give some details on this at the end of Section 3.

**Notation and Simplifying Assumptions**

For the remainder of the paper we will only consider importing and processing data from single tables; which, e.g., correspond to log files at Google in the "protocol buffers" format [29] or result from denormalizing a set of relational tables in a database. We refer to such an instance as *table* or just *the data* which has *columns* (also referred to as *fields*) and *rows* (also referred to as *records*). In order to store protocol buffer records with nested and repeated records (i.e., lists of sub-records), PowerDrill supports a nested relational model, cf. [5]. For ease of exposition, in the following we focus on unstructured / flat records as opposed to records which may, e.g., contain lists.

## 2. BASIC APPROACH

### 2.1 The Power of Full Scans vs. Skipping Data

As mentioned previously, the main advantage column-stores have over traditional row-wise storage, is that only a fraction of the data needs to be accessed when processing typical queries (accessing often only ten or less out of thousands of columns). Another important benefit is that columns compress better and therefore reduce the I/O and main memory usage.

A common characteristic of these system is that they are in most cases highly optimized for efficient full scans of data. In data mining use cases, such as ours, the queries are too diverse for traditional indices being used effectively. The where clause can be free form, allowing to restrict on arbitrary dimensions or even computed values (e.g., all web-searches that contain the term "cat").

As a rule of thumb, even in large database systems if more than a certain, often small percentage of the data is touched, a full scan is performed as opposed to using any indices. The obvious benefits being less random access I/O, simpler, easier to optimize inner loops, and very good cache locality. The latter already easily accounts for a factor of 10 for data which is in memory and when comparing scanning vs. random access: an L2 cache access usually costs less than 1/10th of a main memory access, see, e.g., [12].

The logical next step is to try to combine the benefits of an index data-structure (making it possible to *skip data*) with the power of full scans. This can be achieved by splitting the data into *chunks*[2] during import and providing data-structures to quickly decide which chunks can be skipped at query processing time. On each *active*, i.e., not skipped, chunk a full scan is performed. For our application, partitioning is much more powerfull than traditional indices, since partitions allow indexing by multiple dimensions and enable covering lookups without duplication the data (such costly duplication is, e.g., used by C-Store / Vertica as proposed in [3]). Also, they can take advantage data correlation, e.g., partitioning by country also helps to look for web-searches that contain the term "cat", since mostly Engish speaking countries would contain that word.

This is a generally applicable approach. Consider, e.g., the work of Bast and Weber [7] on a system for interactive literature searches: the underlying data is preprocessed into "blocks" such that for each search only one block needs to be scanned.

The Brighthouse column-store [32] splits the data into "packs" at the import stage. A "knowledge grid" data structure (also built during import) enables skipping such packs when processing queries.

Moerkotte [24] describes "small materialized aggregates" which also allow skipping chunks and is used in Netezza [26]. The approach is rather limitted, since it is only based on comparing with min and max values per chunk. Moerkotte points out that this works well for what he coins as "implicit clustering": the order of dates (like delivery-dates) is often increasing when records are appended over time. E.g., restrictions such as `WHERE delivery_date > "2012-01-01"` can be used to limit the chunks to scan. Compared to the approach described in the present paper, this only covers some very specific cases.

Slezak et al. [32] explicitly avoid complex preprocessing of data during import for speed and data freshness. They give references to other column-stores with similar approaches, which include partitioning of data. The latter is more expensive at import time, but may enable skipping more data.

This is also the approach chosen for the column-store presented in this paper. In the next section we describe a simple partitioning scheme which we apply at import time.

---

[2]aka. blocks, packs, or parts

## 2.2 Partitioning the Data

Most modern database systems provide multiple options for partitioning tables, see [28] for an overview. In our case we perform a *composite range partitioning* [28] to split the data into chunks during the import.

Put simply, the user chooses an ordered set of fields which are used to split the data iteratively into smaller and smaller *chunks*. At the start the data is seen as one large chunk. Successively, the largest chunk is split into two (ideally evenly balanced) chunks. For such a split the chosen fields are considered in the given order. The first field with at least two remaining distinct values is used to essentially do a range split, i.e., a set of ranges are chosen for the field which determine the first and the second chunk. The iteration is stopped once no chunk with more rows than a given threshold, e.g., 50'000, exists. This "heaviest first" splitting generally leads to very evenly distributed chunk sizes (for an analysis in a theoretical framework see [8]).

In practice, a good heuristic is to let a domain expert choose 3–5 fields which amount to a "natural" primary key of the table. As an example, for PowerDrill's own query-logs the date, country, user name, and SQL query may be a good choice. Note that after the partitioning these fields are not treated specially in any way. Principally, any other technique of splitting up the data would work as well.

In Section 6 we give experimental results for how well the described partitioning scheme performs in our production environment on actual user queries.

## 2.3 Basic Data-Structures

In this section we describe the basic layout used to represent the values of an entire column of the underlying table. Since we are discussing a column-store system, each column is stored and can be accessed independently of the others. It is important to note that the order of the data for all columns is the same and corresponds to the (possibly reordered) rows of the original table. In other words, when "synchronoulsy" iterating over all columns, the original rows can be reconstructed. This property of having the same order is important to correctly compute SQL queries for the original table.

Let us now focus on a single column, say `search_string`, and describe the basic data-structures with the help of a concrete example. We assume that the partitioning described in the previous section has been performed and resulted in 3 chunks.[3] In chunk 0 we have the fictious queries ["ebay", "cheap flights", "amazon", ...]. The values of a column are stored in a doubly indirect way using two dictionaries:

1. We introduce a *global-dictionary* which contains all distinct values of the original column, see the leftmost box in Figure 1 for an example. The values are stored in a sorted manner and can be accessed by their integer *rank* also referred to as *global-id* (e.g., 9 → "la redoute"). Conversely, the global-dictionary can also be used to look up the global-id of a given value (e.g., "ebay" → 5).

2. Per chunk we store a *chunk-dictionary* containing $n$ entries, one for each value / global-id occurring in that chunk. The chunk-dictionary can be used to map occurring global-ids to and from integer *chunk-ids*. These

---

[3]In the production system the data is also pre-split into *shards* as a first step. See Section 4 for details.

Data Column: `search_string`

| global-dictionary *dict* | | chunk 0 | | | chunk 1 | | | chunk 2 | | |
|---|---|---|---|---|---|---|---|---|---|---|
| | | chunk-dict | elements | | chunk-dict | elements | | chunk-dict | elements | |
| id | search string | $ch_0.dict$ | | $ch_0.elems$ | $ch_1.dict$ | | $ch_1.elems$ | $ch_2.dict$ | | $ch_2.elems$ |
| 0 | ab in den Urlaub | id | global-id | | id | global-id | | id | global-id | |
| 1 | amazon | 0 | 1 | 3 | 0 | 0 | 5 | 0 | 1 | 0 |
| 2 | cheap tickets | 1 | 2 | 2 | 1 | 1 | 2 | 1 | 3 | 0 |
| 3 | chaussures | 2 | 4 | 0 | 2 | 5 | 1 | 2 | 5 | 2 |
| 4 | cheap flights | 3 | 5 | 4 | 3 | 6 | 4 | 3 | 10 | 4 |
| 5 | ebay | 4 | 12 | 0 | 4 | 7 | 3 | 4 | 11 | 3 |
| 6 | faschingskostüme | | | 0 | 5 | 8 | 0 | | | 4 |
| 7 | immobilienscout | | | 2 | | | 0 | | | 4 |
| 8 | karnevalskostüme | | | 1 | | | 1 | | | 5 |
| 9 | la redoute | | | 3 | | | 5 | | | 2 |
| 10 | pages jaunes | | | 2 | | | 5 | | | 1 |
| 11 | voyages snfc | | | | | | | | | |
| 12 | yellow pages | | | | | | | | | |

Figure 1: Fictious example illustrating the layout of the data in a single data column

are in the range $\{0, \ldots, n-1\}$ and are assigned to the sorted global-ids in an ascending manner, see the three boxes to the right in Figure 1.

The actual values of the column are then represented by a long sequence of chunk-ids per chunk, the so-called *elements*. For example chunk 0 consists of elements [3, 2, ...] which can be dereferenced by lookup in the chunk- and then the global-dictionary to ["ebay", "cheap flights", ...]. With the notation in the figure, to retrieve the element 3 in chunk 0 we find that $dict(ch_0.dict(ch_0.elems[3])) =$ "yellow pages".

There are numerous advantages of this special "double dictionary encoding". It makes it easy to determine which chunks are not active (can be skipped) when processing a query, see next section. Dictionary encodings are a common approach to compress data. Therefore, it is not surprising that our basic data-structures have small memory footprints. See Section 2.5 for experiments which show that this is achieved even for the case where the entire data is treated as one large chunk. These experiments also show that the encoding is particularly well suited for efficiently computing group-by computations over a single field. The second indirection introduced by the chunk-dictionaries has the effect that the elements are comprised of values from a small range of consecutive integers. This is advantageous when further optimizing the memory footprint, see Section 3.

So far we have introduced the general layout of the data structures used. The actual underlying representation of the data depends on the data type and some other factors. It can be as different as a trie (prefix tree) or bit compressed representations tuned to storing numerical values, see Section 3 for more details on this.

For the following two sections we consider simple, "canonical" implementations: For strings like `search_string`s we use sorted arrays for the global-dictionaries. With this, a lookup by global-id is an array access. For determining the rank of a string, one may use binary search. Global-ids and chunk-ids can be stored as 32 bit int values. The chunk-dictionaries are then sorted int arrays of global-ids and the elements are ordered (but not necessarily sorted) int arrays of chunk-ids.

To go from a $ch_i.elem$ value $e_i$ to the underlying value of the data column one can then first determine the global-id of $e_i$ by a simple lookup in the $ch_i.dict$ array followed by a lookup in the $dc.dict$ array. For the other direction, for a given `search_string` $qs$ do a binary search in $dc.dict$ to determine the rank / global-id of $qs$, followed by a binary search in $ch_i.dict$ to determine the element's chunk-id.

### 2.4 How to Evaluate a Query

With the help of the following example query we wish to convey the general idea of how to compute SQL queries on the basic data-structures introduced in the previous section.

```
SELECT search_string, COUNT(*) as c FROM data
 WHERE search_string IN
       ("la redoute", "voyages sncf")
 GROUP BY search_string ORDER BY c DESC LIMIT 10;
```

First it is determined which chunks are *active*, i.e., contain data that matches the where condition. A lookup of the global-ids for the strings in the IN statement gives (9, 11). Now the chunk-dictionaries are checked for these two global-ids with the finding that 9 is not contained in any chunk and 11 only in chunk 2. In other words, there is only one active chunk.

To evaluate the group-by statement per chunk, an integer array `counts` with the same size as the chunk-dictionary, say $n$, is created (here $n = 5$). We then add up the counts in a loop over the elements, i.e., `counts[elements[row]]++` with `row` in $\{0, \ldots, n-1\}$ and `elements[row] IN (4)`, the chunk-id corresponding to the global-id 11. This gives the COUNT(*) values for one chunk. For the overall result a hash-table storing global-id to counts is updated. In the end, the order of all occurring global-ids is determined, the limit applied, and finally the original values looked up in the global-dictionary. Here this gives only one result row: ("chaussures", 3). As we will see in the next section, using a simple array for incrementing counts in the inner loop has important performance benefits.

In the discussion of the example above we introduced the special treatment of IN expressions when deciding which chunks and which rows are active. To cover many typical cases resulting from users "drilling down" into subsets of the data, the system provides special support of the following operators: AND, OR, NOT, IN, NOT IN, =, !=.

### 2.5 Basic Experiments

In this section we present some initial experiments on canonical implementations of the basic data-structures described in Section 2.3. For these experiments (and for all others except the results presented for the production system in Section 6) we measure the performance of full scans. I.e., we do *not* measure the effect of skipping / caching data by making use of the partitioning scheme described in Section 2.2. In effect, for this section we actually do not partition the data at all and instead treat it as one large chunk.

We compare latency and memory usage of the basic data-structures with other data formats and backends: CSV, record-io (binary format based on protocol buffers [29]), and Dremel (as mentioned in the introduction, Dremel is a high performance column-store developed at Google). The formats CSV and record-io are evaluated by backends developed as part of the PowerDrill project.

We have performed the same experiments on each of the successive optimizations presented in Section 3. This way we can nicely pin-point which effects on the memory footprint individual "algorithmic engineering" choices had.

Let us start by introducing the data and SQL queries used for the experiments. For realistic input data we decided to simply use our own logs as source. PowerDrill is used by many teams across Google; in the last years the system has processed more than 60 million queries. We log various facts and metrics for each of these queries. For our experiments we have extracted 5 million rows with the fields `timestamp`, `table_name`, `latency`, and `country`. Since our system is used over a large variety of data-sets, the `table_name` is actually a field with many distinct values (several 100K; note that these logs contain queries against all backends, including Dremel which provides access to many different tables and for which table-names usually include the date). Such fields deserve special attention: they are a lot more resource intensive than others. The experimental results will highlight some of the characteristics of these fields. The field `country` on the other hand of course has only few distinct values, 25 to be concrete. Note, this is the country in which the user's office is located.

On this data we issue three different SQL queries which exhibit distinct properties of the corresponding backends for typical corner cases.

**Query 1:** top 10 countries
```
SELECT country, COUNT(*) as c FROM data
 GROUP BY country ORDER BY c DESC LIMIT 10;
```

**Query 2:** number of queries and overall latency per day
```
SELECT date(timestamp) as date, COUNT(*),
       SUM(latency) FROM data
 GROUP BY date ORDER BY date ASC LIMIT 10;
```

**Query 3:** top 10 table-names
```
SELECT table_name, COUNT(*) as c FROM data
 GROUP BY table_name ORDER BY c DESC LIMIT 10;
```

For each of the backends we ran the queries 5 times on a single commodity machine (2.6 GHz, 8GB RAM, Linux OS) in a single thread, flushing the disk caches before each run. This simulates a realistic streaming situation for CSV, record-io, and Dremel. For the basic data-structures described in 2.3 we assume they fit entirely in memory (note that this is one of the key differences of our system and motivates the extensive optimizations presented in 3). Additionally to the average latency of these 5 runs, we measured the memory usage. I.e., for Dremel and our own data-structures this reflects only the columns present in the individual queries. For CSV and record-io the entire data size is reported, since these are row-wise formats and therefore the entire data needs to be streamed over to compute results.

**Discussion of Results**

Table 1 shows the results for CSV, record-io, Dremel, and our basic data-structures. We will start by discussing the first three systems and then give some explanations for the at first glance surprising outcome concerning our own data-structures.

**Table 1: Comparing CSV, record-io, Dremel, and our own, basic data-structures**

| Query | Latency in ms | | | Memory in MB | | |
|---|---|---|---|---|---|---|
| | 1 | 2 | 3 | 1 | 2 | 3 |
| CSV | 55'099 | 75'207 | 71'778 | 573.3 | 573.3 | 573.3 |
| rec-io | 27'134 | 50'587 | 39'235 | 551.1 | 551.1 | 551.1 |
| Dremel | 7'874 | 18'191 | 48'628 | 27.9 | 60.4 | 90.8 |
| Basic | 20 | 214[4] | 686 | 20.0 | 41.5 | 91.2 |

Comparing the latencies for CSV, record-io, and Dremel, the superiority of Dremel is clearly visible for Query 1 and Query 2 with speed-ups of 2.7–7x. Interestingly, for Query 3 the difference is basically negligible. The overhead of the actual group-by computation for `table_name` dominates everything else. Note that the field has many distinct values, leading to large internal hash-tables; computing the hashes themselves on possibly large strings is already computationally quite expensive.

Notice the increase in latency comparing Queries 1 and 2 across all backends / formats. This stems from the somewhat expensive computation of the function `date(..)` and the additional sum in the group-by statement.

It is important to point out that I/O—i.e., streaming from disk—is not the main bottleneck for any of the queries. Consider for instance that the latencies for Query 3 are relatively similar, but that CSV and record-io need to stream more than 500 MB compared to Dremel which only needs to load 90 MB. Generally speaking, it is reasonable to assume a streaming rate of at least 100 MB/second for pure I/O during these experiments. This gives an overhead of at most 1 second for all of the queries in Dremel and about 5 seconds for CSV and record-io.

Let us now have a closer look at the results for our basic data-structures. The first big surprise is that Query 1 can be computed in 20 milliseconds where the other backends take at least 7 seconds. The reason for this is quite simple. By choice the encoding we use is extremely beneficial for group-by statements over single fields. The inner loop basically boils down to executing the following statement 5 million times: `counts[elements[row]]++`, where `row` is the current row, `elements[row]` gives the chunk-id stored in `country`'s elements array for that row (see also Section 2.3), and `counts` is a relatively small array (with the number of distinct countries as cardinality). This compares favorably to more generic implementations which use hash-tables and can cope with multiple group-by fields[5]. The increase in speed for Queries 2 and 3 is similar.

To obtain these extreme speed-ups, it is crucial to have the data-structures in memory. Consider Query 3: after identifying the top 10 chunk-ids for `table_name` integers (by sorting all chunk-ids by their counts after the inner loop), the original `table_name` string values need to be looked up in the dictionary. If the entire dictionary would need to be loaded into memory only to look up these 10 integers, we would lose a large part of the advantage.

---

[4]On first access, expressions such as `date(timestamp)` are computed and materialized in the datastore as virtual fields. We assume that this has happened before computing Query 2. For more details see below and Section 5.

[5]In PowerDrill multiple group-by fields are combined into one expression which is materialized in the datastore as an additional "virtual" column. Such columns can be used in the same manner as original columns.

But even if we drop this "in memory" assumption, we would still see performance boosts greater than 30x compared to Dremel (assuming an I/O streaming rate of 100 MB/second).

Interestingly, the memory usage of the uncompressed, basic data-structures is about the same as the memory usage of Dremel's compressed format. In other words, in these examples the simple dictionary encoding we use is already as compact as the output of the generic compression algorithm used by Dremel. The self-built encoding has the important advantage of being ready to use (no need to decompress it before usage).

As a final observation, notice that the latency of Query 2 for "Basic" is very small even though the expensive function `date(..)` needs to be computed. This would be an overhead that is basically the same across all backends. Here our column-store profits from an important design decision: all expressions on fields are materialized in the datastore. I.e., on first execution a "virtual field" corresponding to `date(timestamp)` is computed and stored in the same format as all other fields. This not only ensures reuse of data for complex and costly expressions, it also enables using restrictions on such expressions to potentially skip entire chunks (by using the chunk dictionaries), see also Section 5.

## 3. KEY OPTIMIZATIONS

Data-structures with small memory footprints are essential for the overall performance of our system: Compared to other column-stores which can focus on efficiently streaming (possibly from disk) all columns accessed by a query, we heavily rely on as much data being in memory as possible.

Intuitively speaking, a column-store which does full scans, accesses every entry of every column needed to process a query. In a sense, all of the data is "touched" and therefore it is usually affordable to actually stream the data from disk. The streaming incurs an overhead in the same order of magnitude as the actual evaluation of the query.

In contrast, in our case we may only access a fraction of the data represented, e.g., by the global- and chunk-dictionaries. Loading these from disk for each query would lead to a disproportionate overhead. To give a concrete example, loading an entire dictionary for the `table_name` field (from our experiments) from disk, would essentially bring down the performance to the level of streaming all data, i.e., doing full scans. In other words, to really profit from the "basic data-structures" described in Section 2.3, we rely on them being in memory whenever possible.

In this section we describe a selection of step-wise improvements we made to these data-structures. For each of the steps we perform the same experiments as described in Section 2.5; measuring latency and memory usage. This enables us to nicely point out the effect of each of the optimization in the various cases covered.

**Partitioning the Data into Chunks**

As a first step we measure the effect of partitioning the data into chunks with the scheme described in Section 2.2. Recall that for our basic experiments we treated the entire dataset as one large chunk. For the partitioning we use the field order `country, table_name` and we set the threshold for the maximum chunk size to 50'000 rows. This leads to about 150 chunks. The following table shows the memory usage compared to Dremel and our basic data-structures with a single chunk. Here and in all further experiments we do not show the corresponding latencies, since they do not change significantly (the main goal is to reduce the memory footprint).

| Query  | 1     | 2     | 3     |
|--------|-------|-------|-------|
| Dremel | 27.94 | 60.37 | 90.79 |
| Basic  | 20.00 | 41.45 | 91.23 |
| Chunks | 20.07 | 47.99 | 91.32 |

The slight increase in size stems mainly from the many more chunk-dictionaries which are now present. For Query 1 and 3 the increase in size is significantly smaller than for Query 2. The reason being that the corresponding fields `country` (Query 1) and `table_name` (Query 3) are in the field order used for the partitioning, therefore each chunk has relatively few distinct values for these fields and hence small chunk-dictionaries. In contrast Query 2 accesses the `latency` field which has many distinct values for each of the chunks.

**Optimize Encoding of Elements in Columns**

So far we took the simple approach of storing the "elements" (the chunk-ids which describe the actual values for a column in a single chunk) as 32 bit integers. An obvious optimization is to choose a better encoding which depends on the size of the chunk-dictionary. If there is only 1 distinct value, we only need the size of the chunk, say $n$, and the chunk-dictionary to "reconstruct" the original values. This gives a constant $O(1)$ overhead independent of $n$. Similarly, in case there are two distinct values a bit-set suffices; resulting in $\lceil n/8 \rceil$ bytes. We complete the picture by using 1, 2, and 4 bytes per chunk-id for the cases of at most $2^8$, $2^{16}$, and $2^{32}$ distinct values, respectively. Table 2 shows the resulting improvement comparing the memory usage of the elements and chunk-dictionaries only (left side) and the memory usage overall (right side).

**Table 2: Memory usage with optimized columns**

|         | Elements in MB |       |       | Overall in MB |       |       |
|---------|-------|-------|-------|-------|-------|-------|
| Query   | 1     | 2     | 3     | 1     | 2     | 3     |
| Basic   | 20.00 | 40.73 | 24.21 | 20.00 | 41.45 | 91.23 |
| Chunks  | 20.07 | 47.26 | 24.29 | 20.07 | 47.99 | 91.32 |
| OptCols | 0.08  | 22.26 | 14.29 | 0.08  | 22.99 | 81.32 |

The effect of this simple optimization is quite dramatic for Query 1. 80 KB suffice to encode the entire column with 5 million values. The reason for this is that there are only 25 distinct countries in the dataset and the `country` field is the first field in the order used for partitioning the data. Therefore, most of the resulting chunks contain only 1 or two distinct values giving a very compact encoding. I.e., for fields with few distinct values this gives big wins, which is basically obvious by construction.

For the other two queries the savings are also significant, but for the particularly important (and hard) case of a field with many distinct values (`table_name` in Query 3), the overall saving is still relatively small: from 91 MB down to 81 MB. In other words, the global-dictionary dominates the size of the encoding. This gives rise to the next optimization.

**Optimize Global-Dictionaries**

As pointed out, the basic encoding of dictionaries may be very large, they e.g., contain the verbatim strings of all table-names. We made use of two properties when choosing an improved encoding: the dictionaries are sorted (alpha-numerically for strings) and in practice the stored values often have long common prefixes. The requirements for the desired small-footprint data-structure were to support lookups in both directions, i.e., from string value to integer global-id and vice versa. For the former direction, tries (prefix trees) seemed like an ideal choice. See, e.g., [31] for a description of tries. We have implemented a high performance trie data-structure which is built on a handcrafted encoding stored in a large byte array. It relies heavily on finetuned bit manipulations. In order to support efficient lookups from global-id to string without incurring a large memory overhead, the inner nodes are chosen to represent 4 bit parts of the represented strings (as opposed to the more standard choice of characters). On lookup one can afford to iterate over all children of each node along the path to decide into which child to descend. This results in at most 16 operations per node.

In our experiments this trie data-structure drastically reduces the size of the global-dictionary for `table_name` from 67.03 MB down to 3.37 MB. The overall memory usage of Query 3 goes down from 81.32 MB (see "OptCols" in Table 2) to 17.66 MB.

**Generic Compression Algorithm**

Let us now look at the easy and obvious optimization of applying a generic compression algorithm on the encoding. This is basically done by all column-stores, cf. [2]. As mentioned, the resulting excellent compression rates of column-wise storage in comparison to row-wise storage constitutes one of the key advantages of the columnar format. We use Google's own high speed compression algorithm Zippy, externally available as Snappy [33]. In order to put the (so far) successively introduced optimizations into perspective, we apply Zippy to each of the resulting encodings: Basic, Chunks, OptCols, and the optimized global-dictionaries OptDicts. The results are shown in Table 3 with the uncompressed memory usage on the left for comparison and the compressed memory usage on the right.

Table 3: Applying Zippy to the individual encodings

|  | Uncompressed in MB | | | Compressed in MB | | |
|---|---|---|---|---|---|---|
| Query | 1 | 2 | 3 | 1 | 2 | 3 |
| Basic | 20.00 | 41.45 | 91.23 | 3.02 | 17.35 | 17.70 |
| Chunks | 20.07 | 47.99 | 91.32 | 0.28 | 16.34 | 12.19 |
| OptCols | 0.08 | 22.99 | 81.32 | 0.04 | 16.32 | 12.19 |
| OptDicts | 0.08 | 22.98 | 17.66 | 0.04 | 16.32 | 12.40 |

The first point to notice is that Zippy achieves very good compression ratios out of the box and can additionally profit a lot from the partitioning. The reason being that the resulting chunks each contain fewer distinct values. This is particularly visible for Query 1 where Zippy can compress 20 MB down to 0.28 MB after the partitioning was applied. Another very interesting observation is that our further optimizations have virtually no effect on the size of Zippy's output for Queries 2 and 3. Consider the latter where our own encodings reduce the size from 91.23 MB to 17.66 MB and Zippy appears to hit a wall around 12 MB.

Qualitatively speaking, our OptCols and OptDicts encodings exploit similar properties as a generic dictionary based compression algorithm would (which does not use any expensive, entropy based techniques like Huffman codes). But it is still surprising to see that in these cases the final size almost seems like an invariant.

One relevant question to ask is why not just rely on the good compression ratio of Zippy instead of going to the trouble of hand-crafting encodings. The answer is that our encodings are ready to use without any preprocessing and are even designed to allow random access; both to the elements describing the columns and the dictionaries. This is a big advantage, since, as pointed out, only small portions of the data may be accessed.

Nevertheless, the additional gains of 1.4x–2x in compression ratio are significant. To avoid the performance penalty (up to 2x in latency for these experiments), we decided to use a hybrid approach with two "layers" of data-structures held in-memory: uncompressed and compressed. Moving items between these layers or finally evicting them entirely can be done, e.g., with the well-known LRU cache eviction heuristic.

**Reordering Rows**

The next step is to "help" Zippy in compressing the elements (chunk-ids) representing the data columns. An observation made by Johnson et al. [18] and others is that by reordering the rows one may improve the compression ratio. In the setting studied by Johnson et al. reordering the rows does not influence the results of computations. The same of course also holds for SQL queries. They have proven that the problem of finding an optimal reordering is NP-hard, present heuristics, and test them on real-world data.

In the following we will describe how to apply this observation, give a qualitative explanation of why this can improve the compression ratio, and recapitulate a somewhat surprising connection to the well-known traveling sales person problem, stated by Johnson et al. Finally, we will present the simple heuristic we have chosen and give the corresponding results in our experimental setup.

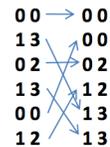

Figure 2: Example of reordering 6 rows with two columns. The reordered version on the right would result in better compression when, e.g., using run-length encoding

Figure 2 gives an example of how reordering can help compression for two columns of elements / chunk-ids. Consider the basic compression algorithm *run-length encoding* (RLE) which replaces consecutive identical values with a counter and the value itself. E.g., the column $0, 0, 0, 1, 1, 1$ would be encoded as $(3, 0), (3, 1)$.

For simplicity we now restrict to the case of columns with only two values as bits 0 and 1. For this case the RLE can be simplified to only storing the counters and not the values themselves. Each time a bit is flipped, a new counter is added, see Figure 3 for an example. This directly gives an interesting characterization of the size of the encoding: for row $r$ the number of bits that differ compared to row $r + 1$ give that row's "contribution" to the total number of counters. Consider the last row in the figure: for columns 1 and 3 two new counters are added on the right.

```
0 1 0         2 3 1
0 1 1           1 1
1 1 0             1
```

Figure 3: Three columns of bits compressed with a simplified RLE

In other words, the encoding size is $\sum_{\text{rows } r} dist_h(r, r+1)$, where $dist_h(r, r + 1)$ gives the Hamming distance between two rows, i.e., the number of differing bits. One can interpret each row of the input as a vector / point in *Hamming space*, the $d$-dimensional space $\{0, 1\}^d$. The distance between two points in Hamming space is the Hamming distance between the two vectors. Each ordering of the original rows corresponds to a path with these points. Figure 4 depicts three rows and the corresponding path in Hamming space. The weights on the edges give the Hamming distance between two points.

```
0 1 0
0 1 1
1 0 0
```
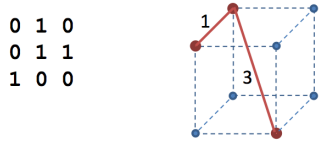

Figure 4: Three rows of bits and the corresponding points in Hamming space and the path given by the ordering

The objective of finding the ordering with the smallest encoding can now be rephrased as finding the path of shortest length. This corresponds to the well-known *travelling-salesperson* problem (in Hamming space). Johnson et al. [18] show that this problem is NP-hard. Earlier, Trevisan [35] has proven that this problem is even hard to approximate within a constant factor for $d > \log n$. Johnson et al. investigate nearest neighbor heuristics and split the data into ranges to deal with the otherwise quadratic runtime (in the number of rows).

For our purposes we chose a very easy to implement heuristic which in practice gives good results: we sort lexicographically by the field order chosen for the partitioning. This has been investigated in more detail, e.g., by Lemire and Kaser [21] or Abadi et al. [3]. The latter describes C-Store's / Vertica's approach of storing columns in multiple sort orders in order to improve the query performance at the cost of storing complete copies of the data.

In our experiments when considering the encoding of the elements and chunk-dictionaries only (without the global-dictionaries), this gives us an improvement of factors 1.2, 1.3, and 2.8 for Queries 1, 2, and 3, respectively. This is compared to compression without reordering.

**Summary**

In Table 4 we show an overview of the step-wise improvements presented in this section, including Dremel for comparison.

Table 4: Summary of the step-wise optimizations presented in Section 3, giving the overall memory usage in MB for each of the variants

| Query | 1 | 2 | 3 |
|---|---|---|---|
| Dremel | 27.94 | 60.37 | 90.79 |
| Basic | 20.00 | 41.45 | 91.23 |
| Chunks | 20.07 | 47.99 | 91.32 |
| OptCols | 0.08 | 22.99 | 81.32 |
| OptDicts | 0.08 | 22.98 | 17.66 |
| Zippy | 0.04 | 16.32 | 12.40 |
| Reorder | 0.03 | 12.13 | 5.63 |

## 4. DISTRIBUTED EXECUTION

**Distributing Data to many Machines**

As described in the introduction, the column store presented in this paper is set up to run in a highly distributed manner. For simplicity of exposition, we have so far only considered a setup on one machine with a relatively small amount of data. To scale up to be able to process billions of rows, the data can be distributed to many machines and processed in parallel.

We are mostly interested in group-by queries and for these it is important that the individual machines do most of the work rather than sending data to a central server. To achieve that, we organize the machines as a computation tree and do the grouping and aggregation on each level of the tree. For this to work, we need to execute the aggregations on multiple levels. This is possible for `SUM`, `MIN`, and `MAX`, i.e. aggregations that can be expressed by associative, binary operations (e.g. `SUM(a, b, c, d) = SUM(SUM(a, b), SUM(c, d))`). Or, if aggregations can be expressed by such associative ones, e.g. `count(*) = SUM(1)` and `AVG(x) = SUM(x) / SUM(1)`. We cannot support count distinct by that. Therefore, we use use an approximative technique described in Section 5.

To illustrate how we execute a group-by-query in parallel, consider this simple query over two tables S1 and S2 (S1, S2 could also represent parts of a table, also referred to as *shards*):

```
SELECT a, SUM(x)
  FROM (S1 UNION ALL S2) GROUP BY a;
```

we rewrite the query to:

```
SELECT a, SUM(x)
  FROM (SELECT a, SUM(x) as x FROM S1 GROUP BY a)
       UNION ALL
       (SELECT a, SUM(x) as x FROM S2 GROUP BY a)
  GROUP BY a;
```

Now the leaf level machines execute the inner select in parallel and send the result to the root of the execution tree. The root executes the outer select. This rewrite can be applied recursively, to support deeper trees. The servers at the leaf level execute "where" clauses and the root executes any "having" statements.

One approach to distribute data may be to distribute the chunks resulting from the partitioning. This is very bad for load-balancing though, since machines that contain active chunks may be heavily loaded while others—which only contain chunks that can be skipped—are idle. A better and actually very common approach is to start by *sharding* (i.e., distributing) the data quasi randomly across the machines. Each shard is on one machine and is then partitioned into chunks as described in Section 2.2. This achieves very good load balancing across machines. It has the additional advantage that the partitioning algorithm can be tuned to work well for a bounded amount of input data. In practice each PowerDrill shard has about 5–7 million rows.

**Reliable Distributed Execution of a Query**

In a cluster of commodity machines that may run arbitrary computations spawned by many users, the load on individual machines may vary dramatically. In effect, individual machines may be completely blocked or may evict processes on demand. One of the greatest challenges of making a distributed column-store production ready is handling these fluctuations well.

An important ingredient to getting this right for our setup was to choose a good replication scheme. A query being distributed to many machines is split up into sub-queries, each being responsible for a certain, distinct part of the data. Instead of sending each sub-query out to only one machine, for reliability we send it out to two machines: the primary and a replica. As soon as one of the two repsonses returns, the sub-query is treated as "answered".

Since the data is loaded dynamically to a machine the first time it receives a query for it, it is important to keep the primary and the replica as much in-sync as possible. The overhead of not having the data fresh in memory on the replica would otherwise be too large and would make the replication scheme inefficient. Therefore, we chose to always compute sub-queries entirely both on the primary as well as on the replica, even if one of them returns early (but we of course do not wait for the slower machine before returning the overall result).

## 5. EXTENSIONS

In this section we briefly touch aspects and extensions of the presented column-store which are important for the productionized system, but for which detailed treatment would be out of scope of this paper.

**Complex Expressions**

As previously mentioned, expressions such as in Query 2 of our experiments `date(timestamp)` are computed once and then materialized in the underlying datastore as "virtual fields". This has important performance benefits. The expression only needs to be computed once, consecutive access can reuse the materialized data. Moreover, the special support for the operators `AND, OR, NOT, IN, NOT IN, =, !=` when deciding whether a chunk can be skipped, also applies to such expressions. For example, for the expression `date(timestamp) IN ('2012-02-29', ...)` the left side of the IN statement is materialized including the chunk-dictionaries. For an individual chunk one can then quickly check whether it is active, i.e., actually contains data for the given dates at all.

In practice, having special support for `AND, OR,...` also ensures that the number of expressions which are stored as virtual fields is relatively small. User-given expressions are split apart by these special operators as far as possible, before actually materializing an expression. Note that a lot of the expressions resulting from typical interactions with the Web UI are actually conjunctions of IN statements, when users are "drilling down" into subsets of the data. What remains after splitting away the special operators are mostly fields and in many cases common "building blocks", i.e., expressions that are re-used frequently.

**Count Distinct**

For many analyses it is important to be able to quickly compute the number of distinct values of a field grouped by another field. As an example, consider counting the number of distinct `table_names` per `country`. This can be a very costly operation for fields with large numbers of distinct values, both with respect to memory and runtime.

We have implemented an approximation algorithm for this problem which was originally introduced in [14]. For an elegant description (and analysis) of the variation which we use, see the first algorithm in [6]. The basic idea of the algorithm is to compute hash values of the field to count distinctly. Of these hashes, the $m$ smallest are determined in a single pass. The threshold $m$ is given by the user and is typically in the order of a couple of thousand. The largest of these $m$ hashes, say $v$, can be used to approximate the count distinct results by $m/v$, assuming that the hash values are normalized to be in $[0, 1]$, cf. [6].

We can profit from a very useful property of both the global- as well as the chunk-dictionaries: the underlying values are sorted ascendingly. This helped in designing and implementing a highly optimized data-structure for collecting and storing the smallest $m$ hash values. The data-structure enables us to support count-distinct queries with comparatively small overhead.

**Other Compression Algorithms**

Additionally to the algorithm used for our experiments in Section 3, we have tested 4 other commodity compression algorithms, including variants provided by the standard libraries ZLIB and LZO. For ZLIB we tested settings with and without additional Huffman coding. The latter gave a perhaps surprising gain of additional 20–30% in experiments, but came with the expected cost of being up to an order of magnitude slower. As result of these experiments we chose a variant of LZO for production, since it gave an about 10% better compression ratio and was up to twice as fast when decompressing compared to Zippy.

**Further Optimizing the Global-Dictionaries**

Even with the trie data-structure described in Section 3, these dictionaries still can be huge in practice. When only few chunks are active for a query, there is actually no need to have the entire dictionary in memory. To this end, we split a dictionary up into sub-dictionaries. One of these representing the most frequent values, each of the others representing values from several chunks combined. When processing a query with few active chunks, only a few of these sub-dictionaries need to be loaded into memory.

To further reduce the situations where a (sub-) dictionary needs to be loaded into memory, we additionally keep Bloom-filters for each dictionary. With these Bloom-filters one can quickly check whether certain values are present in a dictionary at all.

**Improved Cache Heuristics**

It is a known problem in disk-cache / paging algorithms that one-time scans of large files may invalidate the entire cache of pages when the LRU strategy is used, cf. [27]. Similar effects can happen in our system when a single query which accesses a lot of data is processed. We wish to avoid cases where a such a query can negatively impact the caches and therefore the performance of other queries. To this end, we have implemented a more sophisticated cache eviction policy, replacing LRU. We chose an approach similar to the adaptive-replacement-cache presented in [22] and the 2Q algorithm presented in [19].

## 6. PERFORMANCE IN PRODUCTION

Our productionized system is running on well over 1000 machines, the distributed servers altogether using over 4T of main memory.

In a typical use case, a user triggers about 20 SQL queries with a single mouse click in the UI. *On average* these queries process data corresponding to 782 billion cells from the underlying table in 30–40 seconds; under 2 seconds per query. An individual server on average spends less than 70 milliseconds on a sub-query. These measurements and those given below are collected over all queries processed during the last three months of 2011.

The good performance is enabled by the optimized data-structures and by a partitioning of the data which works well for most restrictions used in queries. In practice, it actually turns out that choosing the field-orders for the partitioning scheme is quite straightforward. We never had to go back and finetune the selection of fields and we strongly benefit from correlations in the data.

On average 92.41% of underlying records were skipped and 5.02% served from cached results, leaving only 2.66% to be scanned. Note that additionally to skipping over inactive chunks, we also cache results for chunks which are fully active, i.e., for which the where clause evaluates to true for all rows of the chunk.

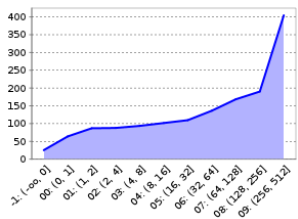

**Figure 5: The average latency of queries in seconds (y-axis) by the amount of data to be loaded from disk in GB (x-axis, as log2-buckets)**

Another interesting question to ask is how many queries could be answered from data-structures which were in memory. On average over 70% of the queries do not need to access any data from disk. They have an average latency of 25 seconds. 96.5% of the queries access only 1 GB or less (cumulative over all servers) of data on disk. The average latency naturally increases with the amount of data which needs to be read from disk into memory, see Figure 5.

## 7. CONCLUSIONS

Column-stores have become very popular in the last years because they allow interactive exploration of large datasets. Take Dremel [23], the column-store used internally at Google. It allows exploration of billions of rows (or, log records) within seconds and its Google-internal usage has almost exploded in the last years. However, there always seems to be a gap between the achievable and desired interactivity—even for "best-of-breed" column-stores running in parallel on thousands of machines (like Dremel).

The column-store presented in this paper pushes the "interactivity limit" out significantly. Our approach is based on two crucial assumptions: (1) The majority of queries are fairly discriminative, similar, and uniform (such as those coming from our Web UI which allows free form queries, but makes it easy to reuse expression "building blocks" and promotes certain interactions such as adding conjunctions of IN statements) and (2) the store has only a few but often explored tables (as opposed to many tables that are not used very often).

Fairly discriminative, similar, and uniform queries enable us to use partitioning on top of the column-wise storage. This helps avoiding full scans of the data. Our deployed scheme based on composite range partitioning of three to five fields works very well in practice: Most queries have restrictions on closely correlated fields which results in skipping of 95% of the input data on average.

Having only a few but frequently accessed tables together with a greatly compactified data-storage allows us to keep most of the data in main memory. Our compaction techniques utilize a dictionary-based data representation together with some more key optimizations (bit-wise element encodings, usage of tries for dictionary-storage, Zippy encoding of values, row-reorderings). Combined, these techniques reduce the data size by up to a factor of 50x.

Summarizing, for usage scenarios like the ones outlined above, our techniques push the limit of interactivity out by one or two orders of magnitude. That is, the amount of data that can be examined interactively increases by a factor of 10–100 compared to traditional column-store technologies.